# DIPOLE SEPTUM MAGNET IN THE FAST KICKER SYSTEM FOR MULTI-AXIS ADVANCED RADIOGRAPHY[*]

L. Wang, S. M. Lund, B. R. Poole, LLNL, Livermore, CA 94550, USA

*Abstract*

Here we present designs for a static septum magnet with two adjacent apertures where ideally one aperture has a uniform dipole field and the other zero field. Two designs are considered. One is a true septum magnet with a thin layer of coils and materials separating the dipole field region from the null field region. During the beam switching process, the intense electron beam will spray across this material septum leading to concerns on beam control, vacuum quality, radiation damage, etc. due to the lost particles. Therefore, another configuration without a material septum is also considered. With this configuration it is more difficult to achieve high field quality near the transition region. Shaped shims are designed to limit the degradation of beam quality (emittance growth). This approach is closely related to a previous septum magnet design with two oppositely oriented dipole field regions presented by the authors [1]. Simulations are performed to obtain the magnetic field profile in both designs. A PIC simulation is used to transport a beam slice consisting of several thousand particles through the magnet to estimate emittance growth in the magnet due to field non-uniformity.

## 1 INTRODUCTION

A linear induction accelerator based X-ray technology can provide time-resolved, 3-D radiography capabilities for a hydrodynamic event. A kicker system, which includes a stripline dipole kicker and a dipole septum magnet, is a key component of this technology [2]. The kicker system cleaves a series of intense electron beam micropulses, and steers the beam into separate beam transport lines to achieve multiple lines of sight on target. The first part of this fast kicker system is a high current stripline dipole kicker that is only capable of imparting a small angular bend to the beam centroid. This is followed by a static field dipole septum magnet that increases the angular separation of the centroids and steers them into separate transport lines.

The ideal "box" geometry of the septum magnet for our application is shown in Figure 1. There are two adjacent apertures where ideally one aperture has a uniform dipole magnetic field and the other zero field, separated by a thin septum in the field transition region. The two beams emerging from the kicker are incident on the magnet a radial distance $d$ from the centerline and are further separated from each other by the dipole field as the beams traverse the axial length $\ell$ of the magnet. Given the beam energy, and the incident and exiting angles of the beam at the dipole septum magnet, the magnetic field needed to provide the desired bend can be calculated. Assuming that the effect of the fringe field is negligible, the required dipole magnetic field $B$ is related to the beam energy $E_b$ and the incident and exiting beam angles $\theta_i$ and $\theta_f$ by

$$B\ell = \frac{mc}{e} \sqrt{\left(\frac{E_b}{mc^2}\right)^2 + 2\left(\frac{E_b}{mc^2}\right)} \left[\sin\theta_f - \sin\theta_i\right] . \quad (1)$$

where $m$ and $e$ are the mass and charge of an electron, and $c$ is the speed of light in vacuum. On traversing the magnet, the beam centroids will move radially a distance

$$\Delta = \ell \frac{\left[\cos\theta_i - \cos\theta_f\right]}{\left[\sin\theta_f - \sin\theta_i\right]} . \quad (2)$$

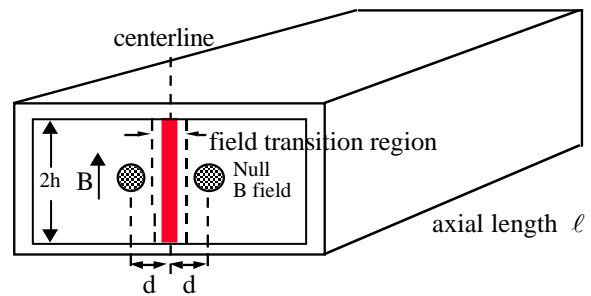

Fig. 1: Ideal box geometry of the dipole septum magnet.

## 2 SEPTUM MAGNET DESIGN

The geometric parameters for the magnet designs are an axial length $\ell = 50\,\text{cm}$, incident and exiting beam angles of $\theta_i = 1°$ and $\theta_f = 22.5°$. This results in a bending field of $B \approx 500\,\text{Gauss}$ for the beam energy of $E_b \approx 20\,\text{MeV}$. The incident beam radius ($r_b$) and centroid separation ($2d$) at the septum magnet are 0.7 cm and 5.6 cm, respectively. The beam centroid will displace a horizontal distance $\Delta \sim 10.4\,\text{cm}$ while traversing the magnet.

The first design considered is a true septum magnet with a thin layer of coils and materials separating the dipole

---

field region from the null field region. Figure 2 shows the 2-D geometry of this septum magnet. The beam goes through the aperture (2*h*) of 10 cm. The amp-turns of each coil for the needed $B$ = 500 Gauss field is $NI \approx Bh/\mu_0 \sim 1990$ amp-turns, where $\mu_0$ is the free space permittivity, and $N$ the number of turns of the coil.

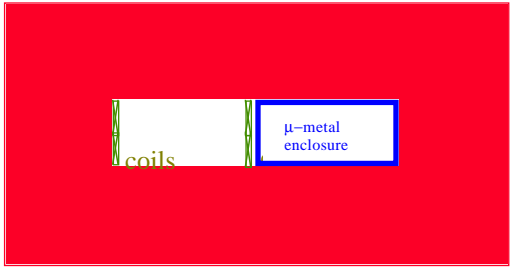

Figure 2: Geometry of a dipole septum magnet with a thin coil separating dipole and null field regions.

During the beam switching process, the intense electron beam will spray across this material septum leading to particle loss and concerns on beam control, vacuum quality, radiation damage, etc. Therefore, another configuration without a material septum is also considered. The schematic of the second design is shown in Figure 3. It essentially consists of two "C" type dipole magnets brought into close radial proximity. With this configuration it is more difficult to achieve high field quality near the field transition region. The rapidity of the transition between the dipole field and null field regions depends critically on $h/d$, the ratio of magnet half-gap $h$ relative to the incident centroid displacement $d$. The ratio $d/r_b$, where $r_b$ is the beam radius, must also be sufficiently large so that the incident beam does not enter the field transition region. Achievable $d$ is limited by the fast kicker technology and the maximum fast kicker to septum magnet drift distance. Thus we need $h$ to be as small as possible in order to obtain good field quality near the field transition region. However, $h$ has to be large enough to allow beam transport through the magnet aperture with negligible scraping over the full range of machine operating parameters. To enhance field linearity and limit the degradation of beam quality (emittance growth), shims are designed and utilized near the field transition region.

## 3 MAGNETIC FIELD SIMULATIONS

The magnetic field in the dipole septum magnet was simulated using the 2-D magnetostatic Poisson code. Only the upper half of the symmetric structure was simulated. Effects due to the iron saturation over the full possible range of coil excitation are negligible. Figure 4 displays the field contours for the dipole septum magnet which has a thin layer of coils in the middle. There is a high quality dipole field on the left side as indicated by the straight field lines.

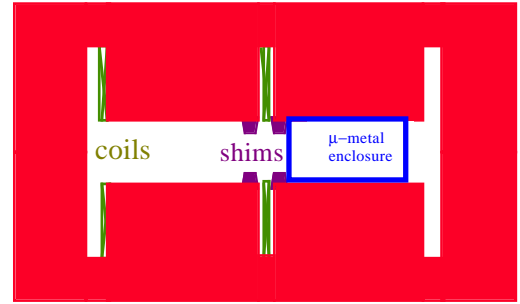

Figure 3: Schematic of a dipole septum magnet which does not have a material septum separating the dipole and the null field regions.

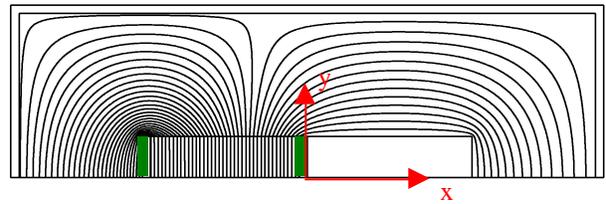

Figure 4: Simulated field lines in the half aperture of the dipole septum magnet with a material septum.

The vertical magnetic field on the midplane (y = 0) of the magnet as a function of *x* is plotted in Figure 5. The predicted magnetic field on the left side is about 500 Gauss as expected. The magnetic field goes from the maximum to a small value when crossing the center septum coil. The electron beam entering the dipole field side of the magnet experiences a uniform field when it traverses the magnet. The simulation result also shows that there is a small leakage field in the null field region. With the use of a thin (3 mm) µ–metal sheet enclosure, the magnitude of the magnetic field leakage is reduced to less than 0.2 Gauss.

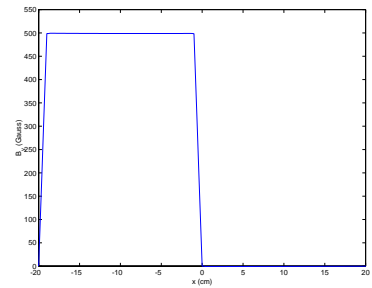

Figure 5: Vertical magnetic field on the midplane of the material septum magnet.

The simulated field contours for the magnet without a material septum separating the dipole field region from the null field region is shown in Figure 6. The semicircles in the figure illustrate the positions of the incident electron beams emerging from the kicker. The curved field lines near the field transition region indicate non-uniform fields. With this configuration, both beams experience non-uniform field which can lead to emittance growth. Compared to the magnet with a material septum, it is more difficult to achieve high field quality near the transition region with this design.

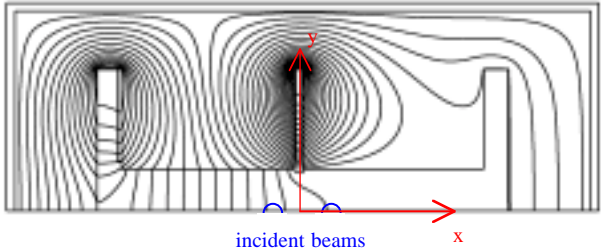

incident beams    x

Figure 6: Simulated field lines in the half aperture of the dipole septum magnet without a material septum.

To enhance field linearity near the transition region, shims of trapezoidal shape are designed. From Figure 7, note that the field lines are straighter near the incident beam position in the dipole field region because of the shims. In addition, a µ−metal sheet enclosure can also be utilized to reduce the leakage field in the null field region. The effect of the shim is further illustrated in Figure 8, where the vertical magnetic field $B_y$ on the midplane ($y=0$) of the septum magnet is plotted. The solid curve represents the magnetic field for the case where no shim nor µ−metal sheet enclosure is used. It takes more than 10 cm for the magnetic field to go from the maximum value to a small number. The leakage field in the null field region is about 80 Gauss at the edge of the incident beam. With the use of the µ−metal sheet enclosure (dashed curve), the field drops to less than 0.1 Gauss at that location. The third curve represents the case where both shims and a µ−metal sheet enclosure are utilized. The curve shows that the shim reduces the undesirable variation in field at the beam edge and centroid for the beam in the dipole field region, allowing the electron beam to experience a more uniform field when it traverses the magnet.

## 4 EMITTANCE GROWTH SIMULATIONS

To determine beam quality for the various magnet designs, a PIC code was utilized to examine the emittance growth in the magnet [3]. A 20 MeV beam slice was transported through the magnetic field region using the 2-D field map for the structure. The results show that there is no emittance growth when the electron beams go through the material dipole septum magnet with a thin layer of coils separating the dipole field region from the null field region. For the design with no material septum, there is no emittance growth when the beam traverses through the null field region and 6% emittance growth when the beam goes through the dipole field region. This emittance growth can be further reduced by refining the design of the shims. These simulations do not take into account the non-ideal effects (vacuum degradation, plasma, etc.) resulting from the particle loss on septum materials.

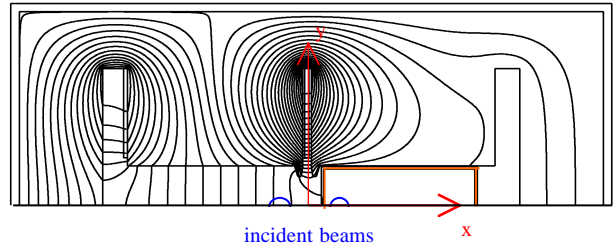

incident beams    x

Figure 7: Simulated field lines in the half aperture of the dipole septum magnet without material septum. Shims are used to make fields more uniform in transition region.

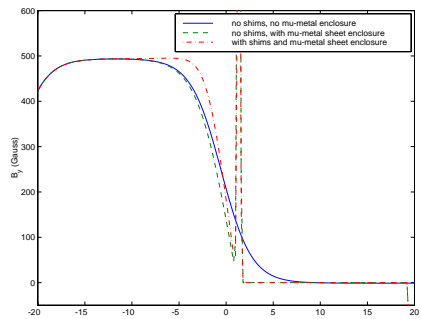

Figure 8: Vertical magnetic field $B_y$ on the midplane ($y=0$) of the septum magnet without a material septum.

## 5 CONCLUSION

Two designs of a dipole septum magnet for the use in multi-axis advanced radiography are presented. One design avoids the use of a material septum. To improve the field quality near the transition region in this design, shaped shims are designed. Simulations are performed to estimate the emittance growth in the magnet for both designs.

## 6 ACKNOWLEDGEMENT

The authors wish to thank G. Caporaso, Y.-J. Chen, and G. Westenskow for helpful technical suggestions.